\documentclass[%
 reprint,
 amsmath,amssymb,
 aps,
]{revtex4-2}

\usepackage{graphicx}
\usepackage{dcolumn}
\usepackage{bm}

\usepackage{cases}
\newcommand{\at}{\alpha_{T/4}}
\begin{document}


\title{Shape-morphing membranes augment the
performance of oscillating foil energy harvesting
turbines}

\author{Ilan M. L. Upfal}
\email[To whom correspondence should be addressed: ]{iupfal@mit.edu}
\author{Yuanhang Zhu}%
 \altaffiliation[Current address: ]{University of California, Riverside}
\author{Eric Handy-Cardenas}
\author{Kenneth Breuer}
\affiliation{%
 Center for Fluid Mechanics, \\ School of Engineering,
Brown University \\
184 Hope St, Providence, RI 02912
}%




\date{\today}

\begin{abstract}
Oscillating foil turbines (OFTs) can be used to produce power from rivers and tides by synchronizing their heaving motion with the strong lift force of vortices shed at their leading edge.
Prior work has shown that compliant membrane OFTs, which passively camber, exhibit enhanced leading edge vortex (LEV) stability and improved lift and power compared with rigid foil OFTs for specific kinematics.
This work seeks to understand a) the performance of compliant membrane OFTs over their full kinematic parameter space and b) separate the roles of membrane camber and extensibility in LEV stabilization.
We characterize the performance of a compliant membrane OFT over a wide range of kinematic parameters through prescribed motion experiments in a free-surface water flume.
The optimal frequency of the compliant membrane OFT is found to be lower than that of a rigid foil OFT due to the enhanced LEV stability of the membrane. 
The lift and power of compliant and inextensible membrane foils are then compared to determine whether camber alone is effective for LEV stabilization or if extensibility plays an important stabilizing role.
The deformation of the compliant membrane OFT is measured using laser imaging.
We observe that the role of extensibility changes for different angles of attack. At low angles of attack, membrane deformation is consistent through the half cycle coinciding with similar performance to the inextensible foil.
At higher angles of attack, the compliant foil has a larger deformation and dynamically decambers corresponding with delayed stall and enhanced lift and power. 
\end{abstract}

\maketitle


\section{Introduction}

While renewable energy deployment has grown dramatically, vast clean energy resources remain unharnessed in river and tidal flows~\cite{doi:10.1289/ehp.115-a590}. These flows are highly predictable, therefore tidal energy and run-of-the-river power may reduce the ancillary service requirements of more variable renewable energy sources such as solar and wind power~\cite{mehmood2012harnessing}.
Horizontal axis rotary turbines (HARTs) are the most mature commercial technology for harnessing tidal power; however, they suffer from high maintenance costs, poor suitability to shallow flows, and high tip speeds that can harm aquatic life. HARTs also significantly decline in performance outside of design conditions and in array configurations~\cite{Khan,garrett2005power,doi:10.1073/pnas.1903680116,Belibassakis2019,Jeffcoate}. 

\subsection{Oscillating Foil Turbines}

An alternate method for harvesting energy from river and tidal is the oscillating foil turbine (OFT)~\cite{wingmill}. The OFT consists of a foil with two degrees of freedom, heaving translation and pitching rotation. The OFT produces power by synchronizing its heaving motion with the strong lift force of vortices shed at their leading edge. 
A significant distance is required between HARTs for the flow speed deficit to recover. The OFT does not require this recovery distance due to the unique wake structure of the turbines~\cite{Franck2021}. Instead of producing a simple velocity deficit wake, OFTs shed strong LEVs which can be exploited or avoided by downstream foils to enhance the system performance of turbine arrays~\cite{ Belibassakis2019,KinseyDumas2012,fenercioglu2015:flowstructuresaround}. OFTs also provide the additional benefits of a rectangular extraction plane more suitable to shallow flows such as rivers and tidal channels and reduced disturbance to aquatic life due to their lower tip speed~\cite{zhu}. 

\subsection{Oscillating Foil Turbine Kinematics}

The kinematics of OFT motion can be characterized by four parameters: pitching amplitude: $\theta_0$, reduced heaving amplitude:  $h^* = h_0/c$, reduced frequency: $f^* = f c / U_\infty$ and phase delay between the heaving and pitching motions: $\phi$. Here, $c$ is the foil chord length, $h_0$ is the heaving amplitude, $f$ is the oscillation frequency and $U_\infty$ is the free stream velocity. The energy harvesting performance of OFTs depends strongly on these parameters and has been studied extensively via simulation, water flume experiments, and field experiments~\cite{kim2017,Kinsey2008,Kinsey2010,su2019:resonantresponseandoptimalenergy,Xiao}.  A phase delay of $\phi = 90^\circ$ has been shown to maximize turbine efficiency~\cite{wingmill,dumas2006eulerian}. The heaving and pitching profiles are commonly sinusoidal, however Su et al. investigated the performance of non-sinusoidal kinematics and found that trapezoidal pitching could improve performance by up to 50\% over sinusoidal pitching~\cite{su2019:resonantresponseandoptimalenergy}. 

Kinsey and Dumas observed formation and shedding of vortices at the leading edge of rigid foil OFTs. Leading edge vortices (LEVs) create a suction force on the foil enhancing power extraction~\cite{Kinsey2008}. LEVs which shed just as the foil reaches the top or bottom of the stroke generally have the greatest strength since they are attached longest. The kinematics with the greatest LEV strength were found to yield highest energy harvesting efficiencies. LEV shedding at the top and bottom of the stroke can also aid the pitch reversal of the foil. An optimal efficiency of 35\% was identified by Kinsey and Dumas at $h^* = 1$, $\theta_o$ = 75$^{\circ}$, and $f^* = 0.15$. This optimal reduced frequency was confirmed and explained by Zhu as corresponding to the most unstable wake mode of the oscillating foil~\cite{zhu_2011}. 

Kinsey and Dumas found the effective angle of attack at mid stroke, $\alpha_{T/4} = \theta_o - \tan^{-1} [\dot h(t = T/4) / U_\infty ] $ where $\dot h(t)$ is the heaving velocity, to be an excellent predictor of OFT performance~\cite{Kinsey2008}. Kim et al. found that OFT efficiency has highest for $30^\circ < \at < 40^\circ$ and $0.09 \leq f^* \leq 0.17$~\cite{kim2017}. Furthermore, the efficiency curves over this frequency range were found to collapse well with respect to $\at$, in agreement with Kinsey and Dumas. Ribeiro et al. focused on the vortex structures in the wake and identified three regimes of operation based on $\at$ and vortex formation: (i) the shear layer regime ($0 < \at \leq 11^\circ $) in which there is no separation at the leading edge, and only small vortices forming in the shear layer behind the foil; (ii)  a leading edge vortex (LEV) regime ($11^\circ < \at <29^\circ$) in which a strong primary LEV is formed, and finally (iii) the leading edge vortex and trailing edge vortex (LEV + TEV) regime ($29^\circ < \at$) in which an additional vortex is formed at the trailing edge of the foil~\cite{ribeiro2021:Wakefoilinteractionsenergy}. As in previous studies, the efficiency of the OFT was found to increase with $\at$ up to $\at \approx 29^\circ$ at which point a maximum efficiency was achieved and the efficiency began to decrease with $\at$. 

\subsection{Compliant and Inextensible Membrane Wings}
The performance of OFTs can be improved by introducing a camber to the foil. The OFTs discussed thus far used rigid foils which must be symmetric and thus cannot have a camber. However, using compliant membrane foils the camber can change between the upstroke and downstroke. Compliant membrane wings are used by flying mammals such as bats and have enhanced lift and a delayed, softer transition to stall~\cite{Song,tzezana,waldman2017camber}. Mathai et al. showed that an OFT utilizing a compliant membrane foil can yield an improvement in lift coefficient of up to 300\% and an improvement in power extraction of up to 160\% by cambering in the flow as well as stabilizing the LEV~\cite{mathai_tzezana_das_breuer_2022}. 

\begin{figure*}
\includegraphics[width = .7\textwidth]{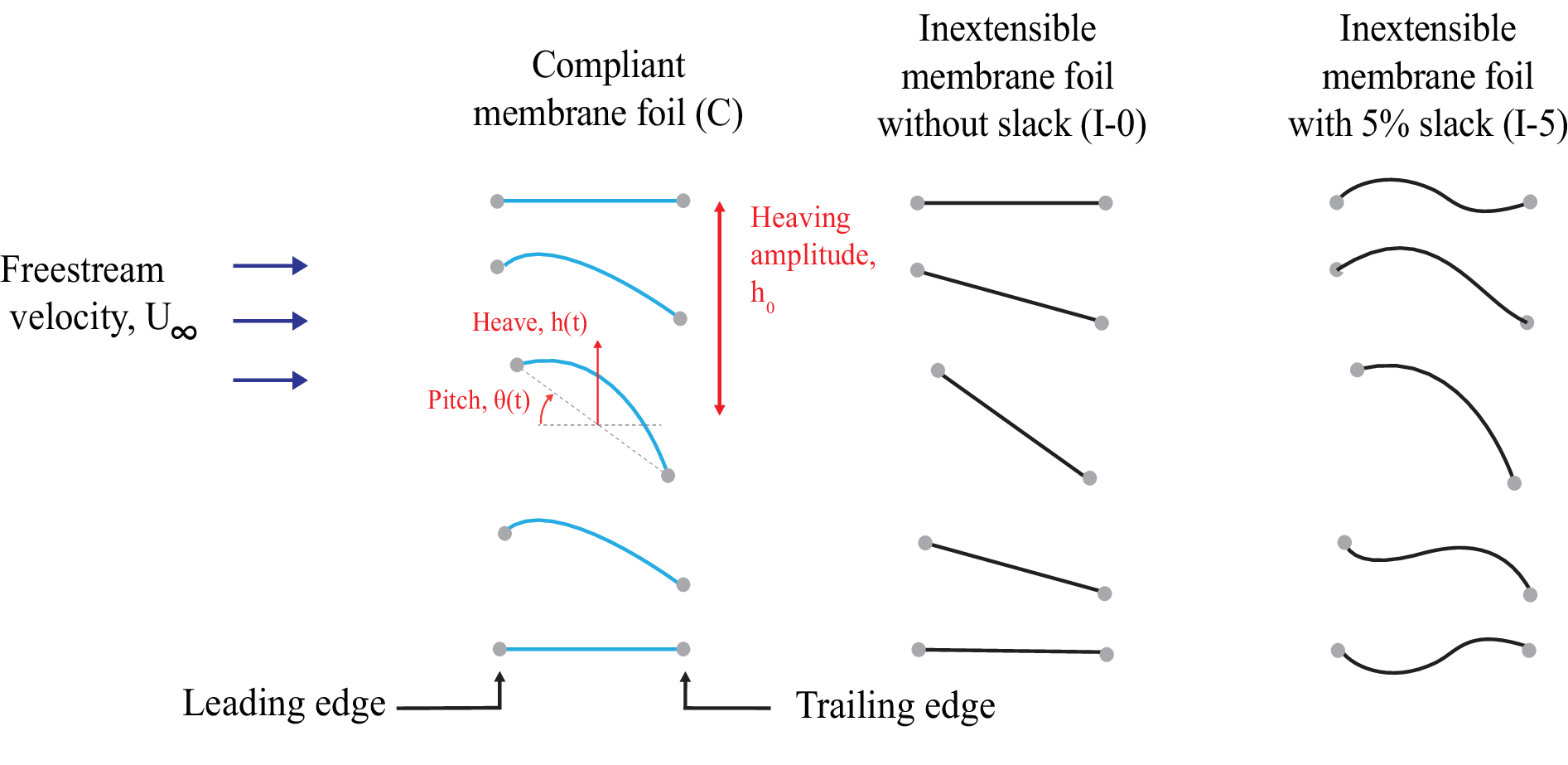}
\caption{Schematic of compliant and inextensible membrane foils in upstroke. The two degrees of freedom, heaving, h(t), and pitching, $\theta$(t), are shown along with the heaving amplitude $h_o$. The C and I-5 foils can be seen dynamically cambering.}
\label{fig:wing_schematic}
\end{figure*}

In the context of energy harvesting, compliant membrane foils offer several appealing features. Since OFTs naturally prefer thin wings with a sharp leading edge \cite{kim2017}, the membranes have ideal geometric qualities.  In addition, their light weight and low bending stiffness provide minimal inertial penalty during pitch reversal. With this in mind the current study aims to extend the work of Mathai et al. with two main goals: first to characterize a wider parameter range, and second to understand the different contributions of camber and extensibility to energy harvesting performance.   In this manuscript we present results of experiments using three OFT configurations (Fig.~\ref{fig:wing_schematic}):  an elastic membrane that can stretch in response to the hydrodynamic forces generated during the cycle (``C''); an inextensible membrane with zero slack that cannot camber (``I-0''), and lastly a membrane with 5\% slack that can adopt a beneficial camber during the upstroke and downstroke (``I-5'').  We measure the membrane shape as well as the power extracted over a range of operating conditions (flow speed, frequency, pitch angle, and heave amplitude) and compare the results with the performance of similar rigid foils.  

\section{Materials and methods}

\subsection{Water Flume Facility}
Experiments were conducted in a free surface water flume at Brown University, (test section width: 0.8 m, depth: 0.53 m, and length = 4.0 m). The freestream velocity, $U_{\infty}$, was set to 28 cm/s for the compliant wing parameter sweep and 32 cm/s for the compliant and inextensible wing experiments, measured using an acoustic doppler velocimeter (Vectrino, Nortek Inc.). 

\begin{figure*}
\includegraphics[width=.6\textwidth]{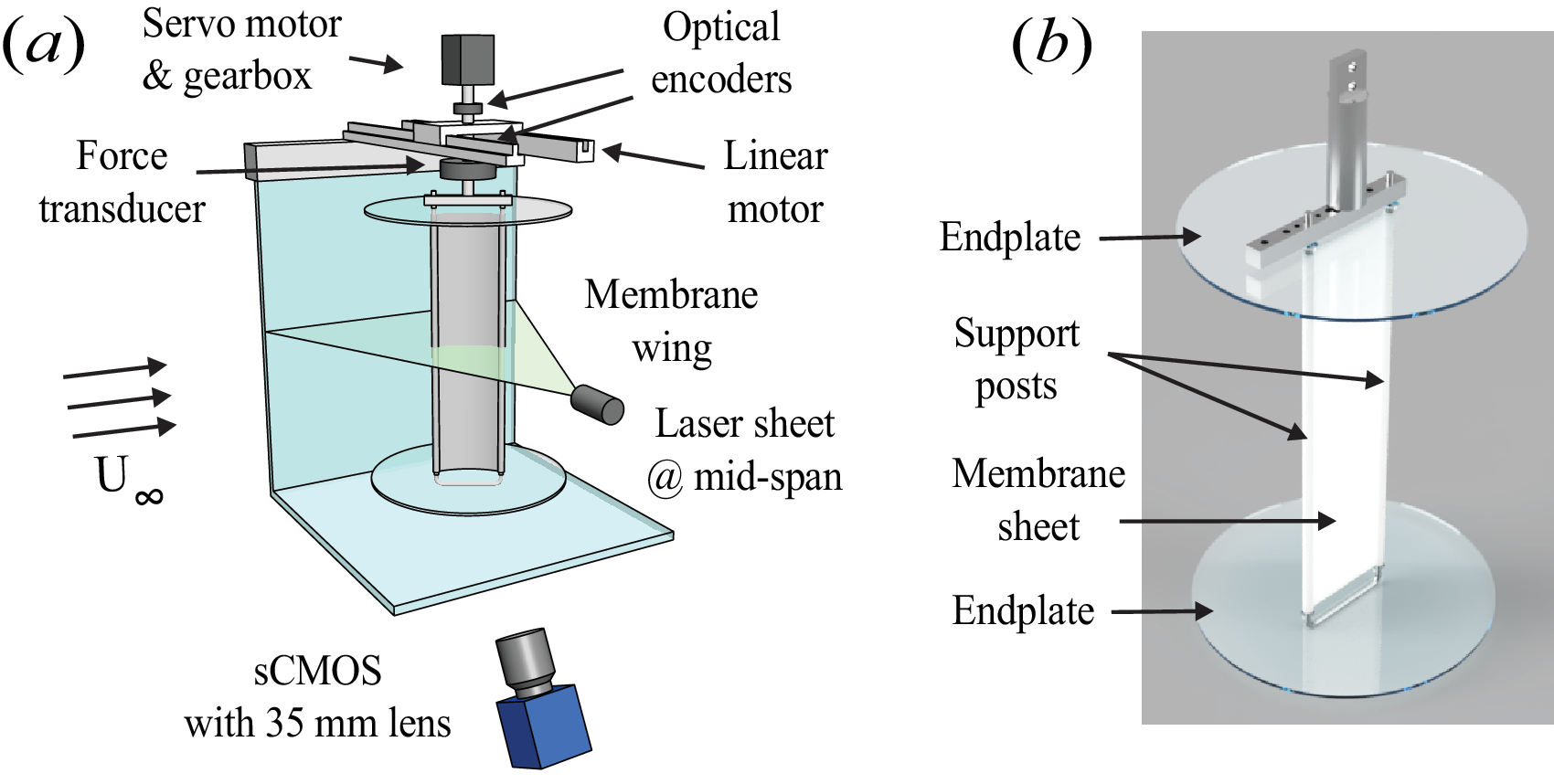}
\caption{a) A schematic of the experimental setup. b) Rendering of membrane foil support frame utilized in this study.}
\label{fig:setup_schematic}
\end{figure*}

The membrane hydrofoil was held from above, supported by a rigid frame (Figure~\ref{fig:setup_schematic}a,b) consisting of a 5 mm diameter steel rod, bent into a U-shape with the legs, 300 mm long and 75 mm apart, defining the wing span, $b$, and chord, $c$ respectively. The base of the U defined the wing tip while the ends of each leg were inserted into a rigid support bar connected to the force transducer and motion carriage positioned above the waterline. The membrane was glued to two steel tubes, outer-diameter: 6.35 mm, using epoxy (Masterbond MS 153).
The tubes slid onto the support frame, thus maintaining a fixed chord length, but allowing free rotation at the leading and trailing edges (LE and TE). Circular end plates were mounted onto the frame above and below the membrane to minimize three-dimensional flow effects.

The frame was mounted on an ATI 9105-TIF-Delta-IP65 six axis force transducer which was used to measure the forces and torques acting on the foil. The entire system was supported by a two-axis heave/pitch system which prescribed the wing kinematics.  A servo motor (Parker SM233AE) controlled the pitching axis motion, while a linear motor (Aerotech BLM-142-A-AC-H-S-5000) drove the heaving motion. Optical encoders were used to record the realized heave and pitch trajectories (US Digital E3-2500-250-IE-D-D-1, and US Digital E3-2500 respectively).

\subsection{Membrane materials}
For the inextensible foil experiments, a thin mylar sheet (100 microns) was glued to the LE and TE tube sections with a slack ratio $s = l/c$ of 1 and 1.05, where $l$ is the membrane sheet length. The compliant membrane material was fabricated in-house by casting a thin silicone membrane sheet using a mass ratio of 50\% Mold Star Series: Platinum Silicone Rubber Part A, 25\% Mold Star Series 16 Fast: Platinum Silicone Rubber Part B, and 25\% Mold Star Series 15 Slow: Platinum Silicone Rubber Part B, with a solvent component of BJB Enterprises TC-5005 Part C added equivalent in mass to 40\% of the silicone mixture. The addition of the solvent reduces the elastic modulus of the membrane to a desired value.  Once mixed thoroughly, the silicone solution was degassed to remove all air bubbles which can act as points of failure in the membrane or change the material properties. The solution was poured onto a clean glass surface and spread using an adjustable wet film applicator (Mitutoyo) at a wet thickness of 750 $\mu$m. The film was allowed to dry thoroughly at room temperature for 36 hours (cured thickness of $h_m = 500 \pm 20$ $\mu$m),  and then laser cut into a rectangular sheet section for use as the test membrane. 

Some membrane samples were also cut into ``dog-bone''-shaped samples and mounted in an uniaxial tensile testing machine (Instron 5942) with which the Young's modulus of the material was determined for the quasi-linear stress-strain region: $1 \leq \lambda \leq 2.4$, where the stretch, $\lambda$, is the ratio of the membrane length, $l$, to its initial length, $l_{i}$.  The Young's modulus of the silicone membrane was determined to be $E = 150$ kPa $\pm$ $5$ kPa. The ratio of elastic stress, $Eh_m$, to the inertial stress, $\rho U_{\infty}^2 c$ yields the nondimensional Aeroelastic number, $Ae = Eh_m / \frac{1}{2}\rho U_\infty^2 c$  which is important in characterizing the strength of the fluid-structure interaction \cite{tzezana,waldman2017camber}.  For the fabricated membrane and the described testing conditions,  $Ae \approx$ 25 and 20 for the compliant parameter sweep and the compliant-inextensible comparison experiments respectively. Once fabricated, the compliant membrane was mounted to the support frame, as described above, so that it had negligible initial stretch, $\lambda_o = 1$.

\subsection{Kinematics}
The first series of experiments conducted covered a broad range of the kinematic operating space of the compliant membrane foil, varying the pitching amplitude in 8 increments of 10$^\circ$ from 15$^\circ$ to 85$^\circ$, the frequency in 8 increments of 0.075 Hz from 0.125 Hz to 0.65 Hz ($f^*$ = 0.037 -  0.195), and heaving amplitude in 4 increments of 0.25c from 0.75c to 1.5c. Sinusoidal kinematics were chosen for these experiments for comparison with related work on rigid foil OFTs. The reduced (nondimensional) heaving amplitude is defined as $h^* = h_o/c$ and the reduced frequency is given by $f^* = fc/U_{\infty}$.

The sinusoidal profiles for heaving, $h(t)$, and pitching, $\theta(t)$, are given by
\begin{equation}
h(t) = h_o \ \cos (2\pi f t), 
\end{equation}
and
\begin{equation}
\theta(t) = \theta_o \ \cos (2\pi f t + \phi), 
\end{equation}
where $h_o$ and $\theta_o$ are the heaving and pitching amplitudes respectively; $f$ is the frequency of oscillation, and $\phi$ is the phase shift between pitching and heaving cycles which was held at 90$^\circ$ in all trials. 

A second series of experiments compared the performance of compliant and inextensible foils. Two foils with inextensible membranes were tested ($s = 1$ and 1.05), and compared with the elastic membrane ($\lambda_o = 1, Ae \approx 20$).  All wings were tested at a single frequency, $f^* = 0.04$ and constant heaving amplitude: $h_o/c = 1.2$; the pitching amplitude varied between $\theta_o = 18^\circ - 57^\circ$. These combinations were chosen so that the effective angle of attack at mid-stroke, $\at$, varied between $15^\circ$ and $45^\circ$. Prior work on the compliant membrane OFT has studied $0 < \at < 15^\circ$. By studying a range of $\at$ spanning from the domain of prior work to higher $\at$, we hope to both validate our results with prior work and gain an understanding of compliant and inextensible membrane OFT performance with increased $\at$. Rigid OFTs obtain a maximum efficiency at $\at = 30^\circ - 40^\circ$, therefore it is of interest to see how the compliant membrane OFT performs under these kinematics.


Non-sinusoidal kinematics (trapezoidal pitching and triangular heaving) were chosen for these experiments for ease of comparison with prior work on compliant membrane OFTs. \cite{mathai_tzezana_das_breuer_2022} Following Su et al., the non-sinusoidal kinematics are conveniently defined by a single parameter, $\beta$, which modulates a cosine curve from trapezoidal (for positive values of $\beta$) to triangular (for negative values of $\beta$)~\cite{su2019:resonantresponseandoptimalenergy}. Note that the equations used by Su et al. have been phase-adjusted for consistency with the rest of the present study:
\begin{numcases}{h(t) =}
    \frac{h_o \sin^{-1} (- \beta \cos (2 \pi f t ))}{\sin^{-1}(- \beta)}  
    & $- 1 \leq \beta \leq 0$ \\
    h_o \cos (2 \pi f t) & $\beta = 0$ \\
    \frac{h_o \tanh [\beta \cos (2 \pi f t)]}{\tanh (\beta)} 
    & $0 < \beta $
\end{numcases} 
and
\begin{numcases}
  {\theta(t) =}
    \frac{\theta_o \sin^{-1} [- \beta \cos (2 \pi f t + \phi)] }{\sin^{-1}(- \beta)}  & $- 1 \leq \beta \leq 0$  \\
      \theta_o \cos (2 \pi f t + \phi) & $\beta = 0$ \\
     \frac{\theta_o \tanh
      [\beta \cos (2 \pi f t + \phi)]}
      {\tanh (\beta)} & 
      $0 < \beta $. 
  \end{numcases}

\subsection{Measurement procedures}

At each operating condition data was acquired over 30 cycles with the first and last three cycles discarded to eliminate the startup and stopping transients. Two metrics are used to characterize the energy harvesting performance of the membrane hydrofoil turbine: the coefficient of power, $C_p$, and the Betz efficiency, $\eta$. The coefficient of power, $C_p$, is calculated from the sum of the cycle-averaged coefficients of heaving and pitching power, normalized by the dynamic pressure and the wing area:
\begin{equation}
C_p= \frac{<F \cdot \dot{h}> + <\tau \cdot \dot{\theta}>}{\frac{1}{2}\rho U_\infty^3 bc}.
\end{equation}
Here, $F$ is the lift force (perpendicular to the flow), and $\tau$ is the pitching moment. $\dot{h}$ and $\dot{\theta}$ are the heaving and pitching velocities, respectively.
The Betz efficiency, $\eta$, is the power normalized by the \emph{swept area} of the oscillating foil:
\begin{equation}
\eta =  \frac{<F \cdot \dot{h}> + <\tau \cdot \dot{\theta}>}{\frac{1}{2}\rho U_\infty^3 A_s} , 
\end{equation}
where $A_s$ is the swept area. Note that $A_s$ is generally not the same as $b h_o$,  due to the pitch angle of the foil.

\section{Results and discussion}

\subsection{\label{sec:level2}Kinematic parameter sweep}
\begin{figure*}
\centerline{\includegraphics[width=\textwidth]{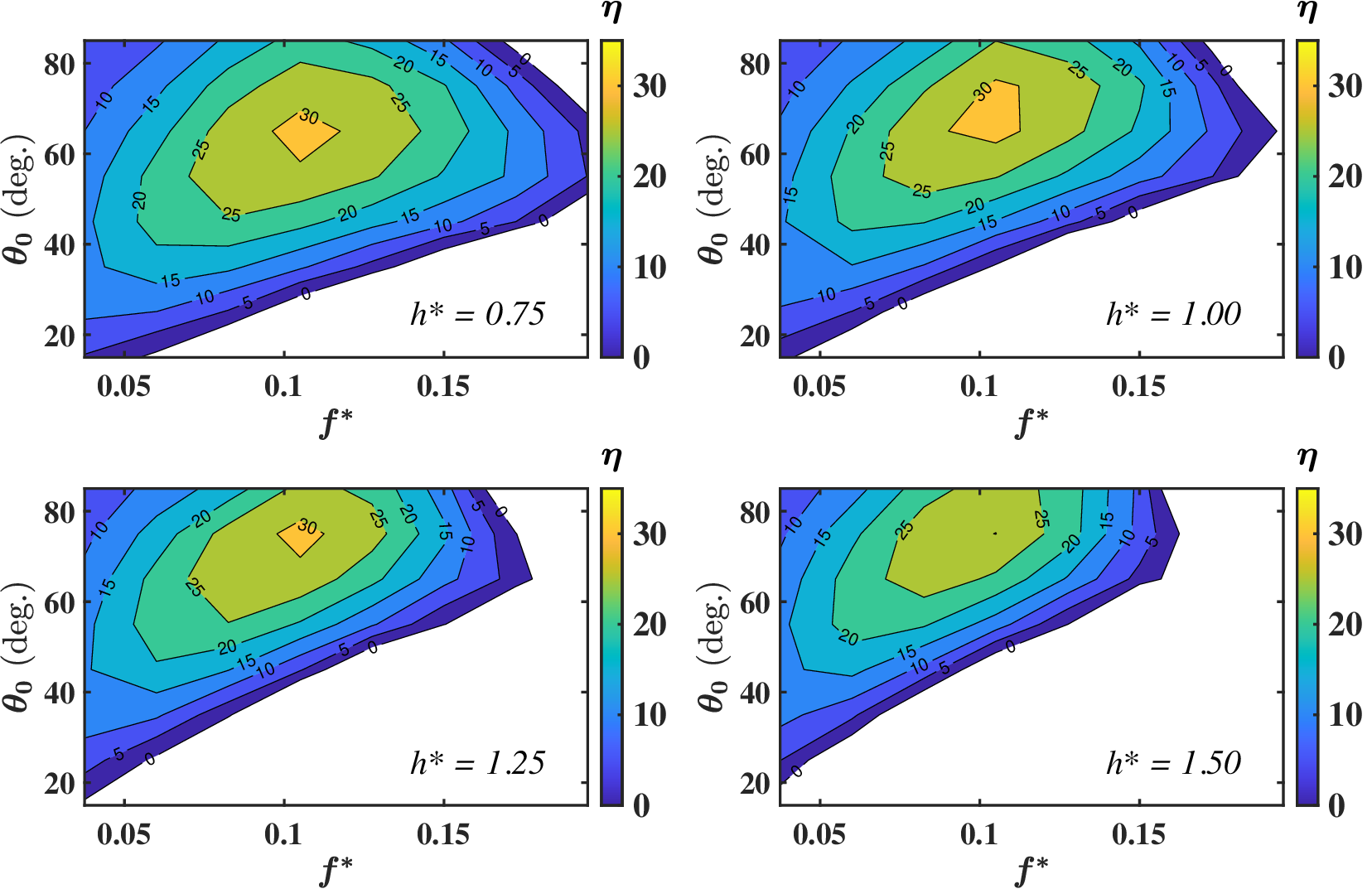}}
\caption{Efficiency, $\eta$, of compliant membrane OFT with respect to reduced heaving amplitude, $h^*$, pitching amplitude, $\theta_0$, and reduced frequency, $f^*$.}
\label{fig:efficiency_map}
\end{figure*}

The energy harvesting performance of the compliant membrane OFT is evaluated first using the Betz efficiency, $\eta$, and second using the power coefficient, $C_p$. 
Figure~\ref{fig:efficiency_map} shows the Betz efficiency, $\eta$, plotted with respect to the reduced frequency, $f^*$, and pitching amplitude, $\theta_o$, for four different heave amplitudes, $h^*_o$.  The efficiency is convex with respect to all parameters tested indicating that a true optimum was found. An optimal efficiency of $31.7 \pm 0.8 \%$ occurs at $h = 1.00$, $f^* = 0.11$, and $\theta_o = 65^\circ$ (Figure \ref{fig:efficiency_map}, top right panel).  The map of the Betz efficiency for the compliant membrane OFT (Figure~\ref{fig:efficiency_map}) closely resembles analogous maps for rigid foils~\cite{Kinsey2008,kim2017}, although key differences exist. 
Notably, the optimal efficiency of the compliant membrane OFT occurs at a significantly lower reduced frequency, $f^* = 0.11$,  than has been found for a rigid foil OFT,  $f^* = 0.15$~\cite{Kinsey2008,kim2017}.
At the large angles of attack of the optimum ($30^\circ < \at < 40^\circ$) the OFT is operating in the dynamic stall regime \cite{ribeiro2021:Wakefoilinteractionsenergy} in which a leading edge vortex (LEV) forms on the suction surface of the wing \cite{Kinsey2008}, and the performance of the OFT is strongly dependent on the synchronization of the growth and shedding of the LEV with the pitch reversal of the wing \cite{Kinsey2008}.  In general, the LEV increases the lift force, enhancing the foil efficiency.  However, at low frequencies, the LEV sheds before the pitch reversal takes place, while at high frequencies, the pitch reversal occurs before the LEV has had time to act, resulting in a drop in the lift force which depresses the efficiency. At a ``sweet spot'', in this case $f^* \sim 0.1$, the contributions of LEV growth and shedding are balanced, making an optimal contribution to the OFT efficiency. Since prior studies have shown that optimal frequency coincides with the synchronization of vortex shedding and pitch reversal, we expect to see a lower optimal frequency for the compliant membrane OFT because the leading edge vortex is more stable on the compliant membrane wing compared to the rigid wing~\cite{mathai_tzezana_das_breuer_2022,Song}. Therefore, the OFT must oscillate slower in order to synchronize with the delayed vortex shedding.


\begin{figure*}
\includegraphics[width=\textwidth]{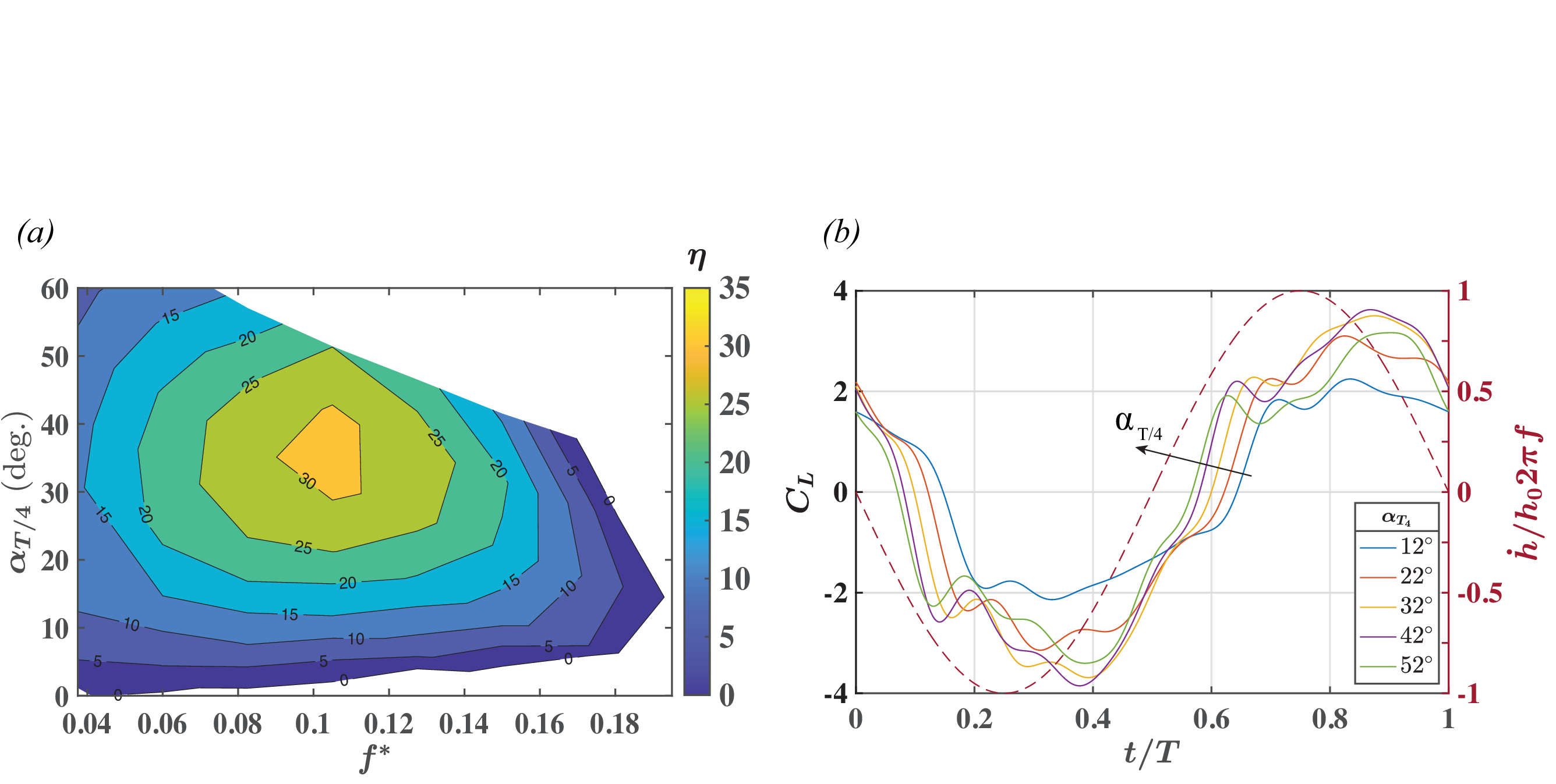}
\caption{a) Efficiency map of compliant membrane OFT plotted with respect to reduced frequency and the effective angle at mid stroke, $\alpha_{T/4}$. Heaving amplitude is held constant at $h^* = 1$. For low $\alpha_{T/4}$, the power coefficient only depends on $\alpha_{T/4}$ and a frequency dependence emerges as $\alpha_{T/4}$ increases. b) Lift profiles of compliant hydrofoil over one cycle as $\alpha_{T/4}$ increases and reduced frequency is fixed at $f^* = 0.11$.}
\label{fig:alpha_lift}
\end{figure*}

Another common trend between the rigid and compliant foil OFT efficiency maps is the steep gradient up from the feathering limit which coincides with the $\alpha_{T/4}$ gradient. To elucidate the relationship between energy harvesting performance and the effective angle of attack at mid-stroke, $\alpha_{T/4}$, the efficiency, $\eta$,  was re-plotted with respect to the reduced frequency, $f^*$, and $\alpha_{T/4}$ (Figure~\ref{fig:alpha_lift}a) for the $h^* = 1.00$ case. Shown this way, it is clear that for $\alpha_{T/4}$ below $\sim 10^{\circ}$, the efficiency is a strong function of $\alpha_{T/4}$ and is largely independent of $f^*$. For $\alpha_{T/4}$ greater than 10$^{\circ}$, the efficiency becomes increasingly frequency dependent. 

For $\at < 10^\circ$, we do not expect a leading edge vortex to be generated on the foil~\cite{ribeiro2021:Wakefoilinteractionsenergy}. For these cases, increasing $\alpha_{T/4}$ increases the lift force, $L$, 
\cite{anderson2017:Fundamentalsaerodynamics} which in turn increases the efficiency, $\eta$. 
For $\alpha_{T/4} > 10^\circ$ we begin to see a convex relationship between efficiency and frequency, with a maximum efficiency at approximately $f^*  = 0.1$. This frequency dependence coincides with a transition to the dynamic stall regime evident in the force measurements and consistent with results of Ribeiro et al.~\cite{ribeiro2021:Wakefoilinteractionsenergy}. This frequency dependence will be further discussed shortly, but we first focus on 
the $\at$ dependence of $\eta$.


More insight into the dependence of performance on $\alpha_{T/4}$ can be gained from the lift-vs-time profiles for a range of $\alpha_{T/4}$. A sample of $C_L$ vs $t/T$ is shown in Figure~\ref{fig:alpha_lift}b for cases of $f^* = 0.11$. The energy harvested by the heaving of the foil is simply the integral of the product of the lift force, $C_L$, and the heaving velocity, $\dot h$, plotted in dashed red (Figure~\ref{fig:alpha_lift}b). In the $\at = 32^\circ$ and $\at = 42^\circ$ cases shown in Figure~\ref{fig:alpha_lift}b the instantaneous power, $|F(t) \cdot \dot h(t)|$, is maximized due to the large overlap between the force and heave velocity profiles relative to the other cases. The compliant membrane OFT was found to have the same optimal $\at$ range as rigid foil OFTs based on previous work~\cite{Kinsey2008}.

In the $\at = 12^\circ$ case (blue line) we observe a delay between the pitch reversal, which occurs at $t/T = 0.5$, and the sign reversal of the lift force, which only occurs at $t/T \approx 0.65$. Prior work \cite{Song,waldman2017camber} and laser deformation measurements (presented in the following section of this paper) reveal that as a compliant membrane wing rotates from a positive to negative angle of attack, the wing retains its positive camber through the pitch reversal, suddenly ``snapping through'' only after the wing has reached a threshold angle of attack. We believe this delayed snap-through behavior to be the cause for the delay in the change of sign of the lift force following pitch reversal. As $\at$ increases in Figure~\ref{fig:alpha_lift}b, this delay reduces and the point at which the lift force changes sign from negative to positive following the pitch reversal occurs earlier. In order to change $\at$ between trials, only the pitching amplitude, $\theta_o$, was varied. An increase in $\theta_o$ coincides with an increase in the speed of pitch reversal such that the wing achieves the threshold snap-through angle earlier in the cycle. This trend is confirmed in the laser imaging of the membrane shape (presented in the following section).

In the $\at = 12^\circ$ case, the foil achieves $C_L \approx 2$ at $t/T \approx 0.7$ and remains close to this value until the following pitch reversal despite small oscillations in the lift force. Similar oscillations were observed by Mathai et al. \cite{mathai_tzezana_das_breuer_2022} and found to coincide with oscillations in the membrane deformation. Such oscillations are observed in the laser measurements presented in the following section.

In the higher $\alpha_{T/4}$ lift profiles presented in Figure~\ref{fig:alpha_lift}b, the foil transiently achieves $C_L > 3$. We observe that the lift force grows quickly after the pitch reversal at $t/T = 0.5$.  This high transient force is associated with the rapid growth of a leading edge vortex in this $\at$ regime~\cite{Ribeiro_2021_PhysRevFluids}. At the highest value of $\at$ tested, this trend is broken and the wing experiences a slower growth in lift force, only exceeding $C_L = 2$ at $t/T \approx 0.8$.  At this very high pitch angle and high pitching velocity, which is well beyond the known rigid foil optimum, this effect might be due to the LEV detaching too early, before it has had an opportunity to sufficiently grow. However, more detailed examination of the flow field, which is beyond the scope of the current work, will be needed to fully explain this observation.

For the three highest $\alpha_{T/4}$ cases (Fig~\ref{fig:alpha_lift}b orange, purple and green lines), the maximum $C_L$ occurs near $t/T$ = 0.9, while for the two lower $\alpha_{T/4}$ cases, the maximum occurs earlier after which the lift drops off until the foil turns over. The $\at = 42^\circ$ case has the highest initial peak amplitude and overall maximum $C_L$ amplitude. Interestingly, for the higher $\alpha_{T/4}$ cases, the coefficient of lift continues to increase until the foil begins to turn over. As mentioned earlier, prior work has identified the synchronization of LEV shedding at $t/T = 0, 0.5$ to be an important factor in optimal kinematics. These results support this hypothesis, since  the lift drops off just before pitch reversal in the $\at = 32^\circ, 42^\circ$ cases (highest performing) suggesting that LEV shedding is occurring at that point in the cycle~\cite{Kinsey2008}. 

\begin{figure*}
\centerline{\includegraphics[width=.9\textwidth]{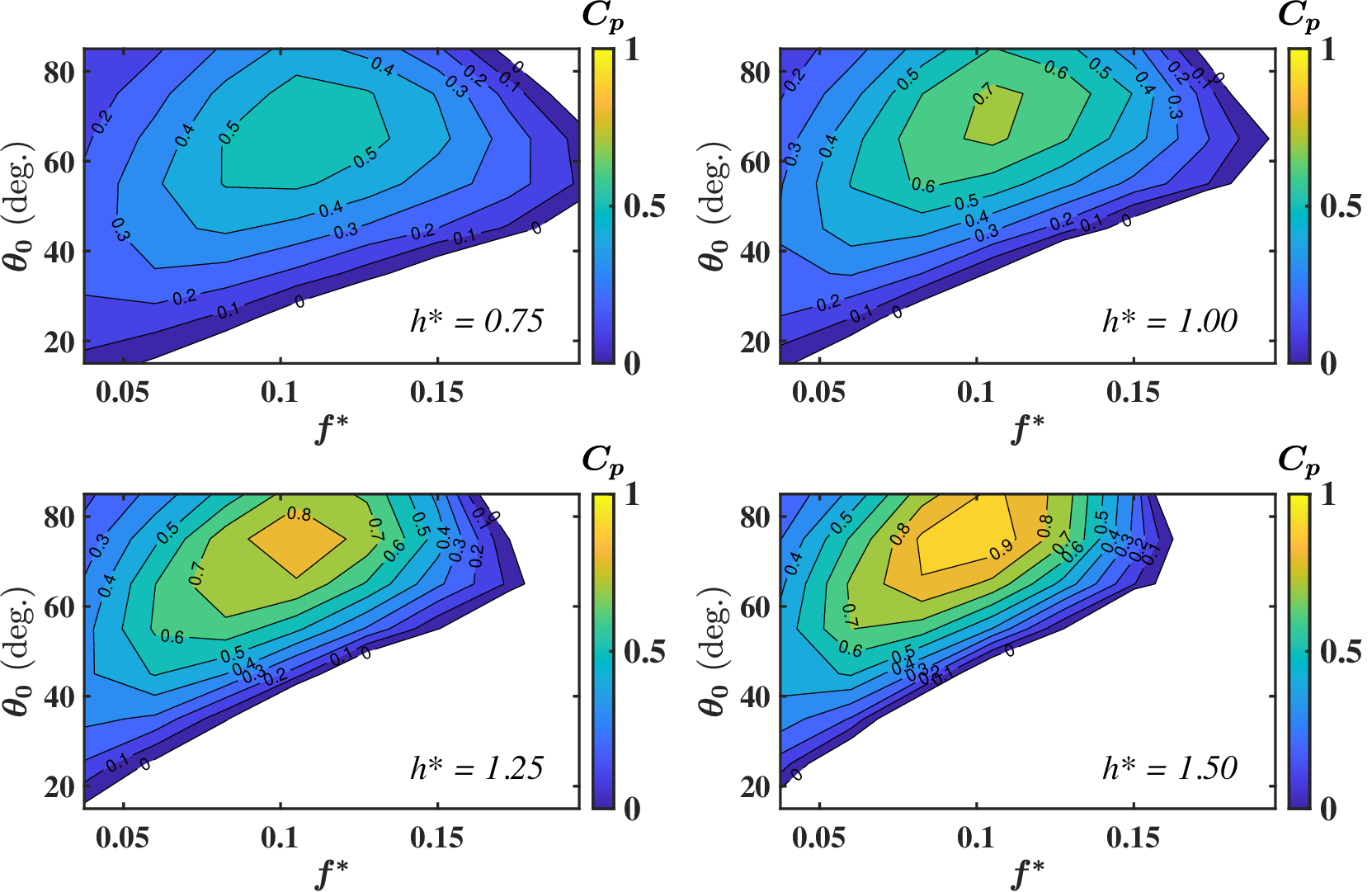}}
\caption{Power coefficient, $C_p$, of the compliant membrane OFT with respect to reduced heaving amplitude, $h^*$, pitching amplitude, $\theta_0$, and reduced frequency, $f^*$.}
\label{fig:power_map_coarse}
\end{figure*}

The coefficient of power, $C_p$, (Figure~\ref{fig:power_map_coarse}) was found to be convex in $f^*$ and $\theta_o$ but increased monotonically with $h^*$. A maximal power coefficient of $0.98 \pm 0.03$  was achieved at $h^* = 1.50$, $f^* = 0.105$, and $\theta_o = 75^\circ$ (Figure \ref{fig:power_map_coarse}, bottom right panel). The power map of the OFT looks similar to the efficiency map in terms of general shape in Figures~\ref{fig:power_map_coarse} and~\ref{fig:efficiency_map} respectively. 
Note that the optimal frequency decreases slightly as the heave amplitude increases. As $h^*$ increases, the heave velocity, effective angle of attack and effective leading edge velocity all increase, resulting in a faster LEV growth. In order to account for this, and to time the pitch reversal with the LEV separation (as discussed in the previous section) the optimal frequency is reduced.
While the efficiency is maximized at $h^* = 1$, the coefficient of power continues to increase with heaving amplitude up to the highest value tested. 
This indicates that the most power is generated from a given cross section of the flow at $h^* = 1$ but more power can be generated per foil by simply oscillating over a larger cross section of the flow. In a practical setting where sea or river bed space is abundant, minimizing equipment costs is most important and maximizing power generation per foil may be the primary design consideration.




We should note that at larger heave values, we may expect that blockage effects in the test section will enhance the OFT performance  \cite{Franck2021,fenercioglu2016:effectofsidewalls,su2019:confinementeffectsonenergyharvesting,ross2020:experimentalassessmentanalytical}.  However in this study, the minimum gap between the foil and the walls is 3.9 chord lengths when $h^* = 1.50$ therefore while some improvement is expected, we expect it to be small.  In addition, the comparisons between different foils discussed in this work, as well as comparisons with the results of Kim et al. \cite{kim2017}, which were performed in the same facility, are consistent.

\subsection{\label{sec:level2}Comparison of Compliant and Inextensible Membrane Hydrofoils}

\subsubsection{Membrane kinematics}
\begin{figure*}
\centerline{\includegraphics[width=\textwidth]{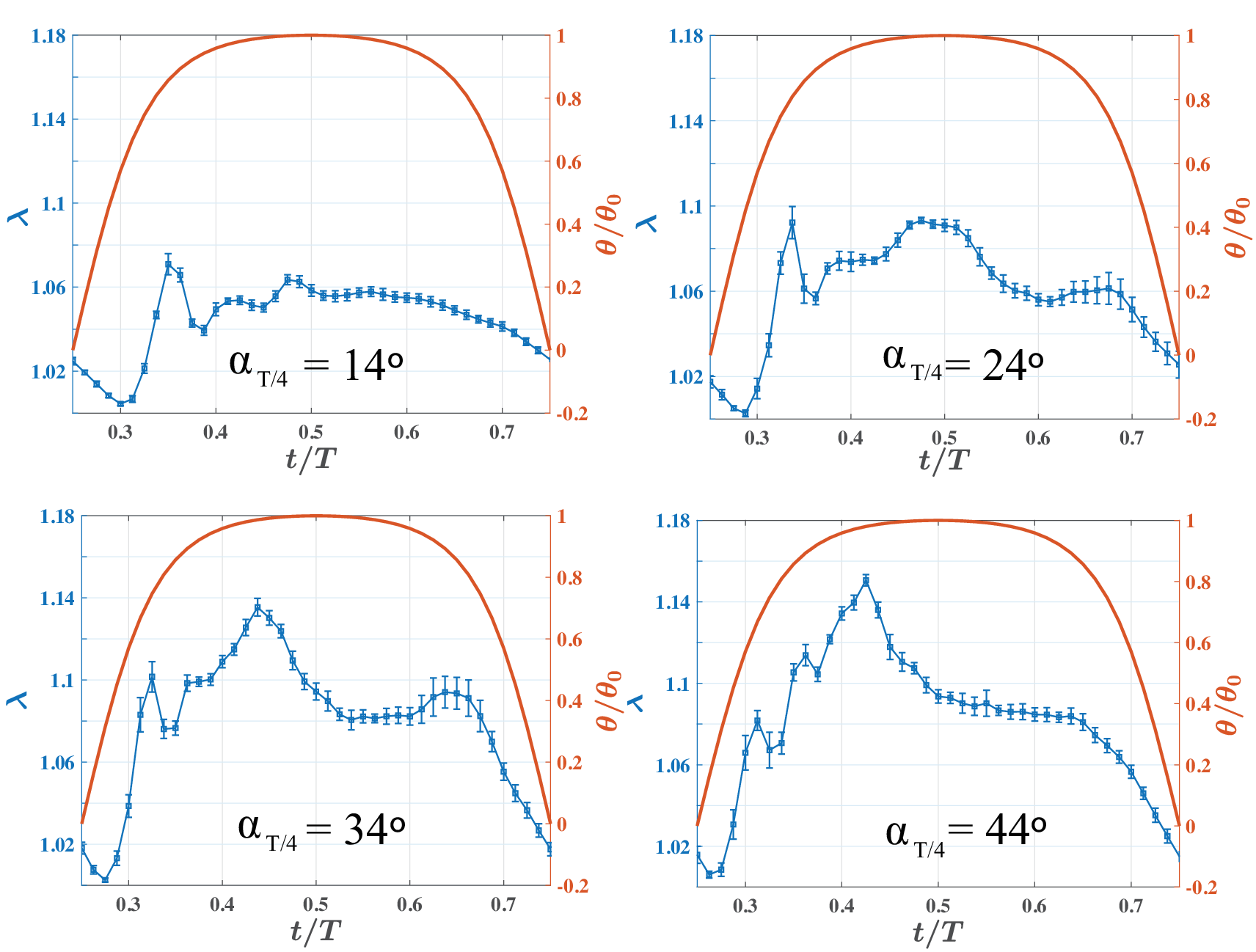}}
\caption{Compliant membrane stretch (in blue) over one half-cycle, for four cases of the effective angle of attack at mid-stroke, $\alpha_{T/4}$ performed using $f^* = 0.04$ and $\theta_o = 18^\circ, 31^\circ, 44^\circ, 57^\circ$. The pitching profile, $\theta / \theta_o$, is shown (in red) for reference.}
\label{fig:stretch}
\end{figure*}

The vital difference between previous studies of oscillating foil turbines using a rigid foil and the current work is the ability of the membrane foil to (i) adopt camber during the power stroke, and (ii) stretch in response to the hydrodynamic forces acting on the wing. The dynamic stretch, $\lambda$, of the compliant membrane is shown in Figure~\ref{fig:stretch}, plotted over one half cycle ($t/T = 0.5 \ldots 1.0$) for four values of $\alpha_{T/4}$. The pitching profile, $\theta / \theta_o$, is superimposed for reference. The initial stretch is one, and as expected, the membrane stretches due to the fluid loading and the stretch increases with $\at$. In all cases, the measurements reveal a local peak in the membrane stretch between $t/T = 0.55$ and 0.6, followed by a transient oscillation. This initial loading shock and oscillation is associated with the elastic vibration of the membrane when it encounters hydrodynamic loading following pitch reversal. 

The membrane stretch behavior differs significantly as the effective angle of attack at mid-stroke, $\alpha_{T/4}$ increases from $14^\circ$ to $44^\circ$ and two distinct behaviors can be identified. In the smallest pitching angle case, despite the initial transient, the membrane stretch remains centered around a constant value of $\lambda \approx 1.05$. At the highest pitching amplitude case of $\alpha_{T/4} = 44^\circ$, there is a steady growth in the stretch until it reaches 16\% at $t/T \approx$ 0.7, followed by a quick decline to a constant value of about 10\%. In summary, we observe the overlay of a vibrational excitation of the elastic membrane with a second phenomena which will be explored further as it relates to differences in leading edge vortex formation and shedding between the $\alpha_{T/4}$ cases.

\subsubsection{Power and lift comparison}

\begin{figure}
\begin{center}
\centerline{\includegraphics[width=.46\textwidth]{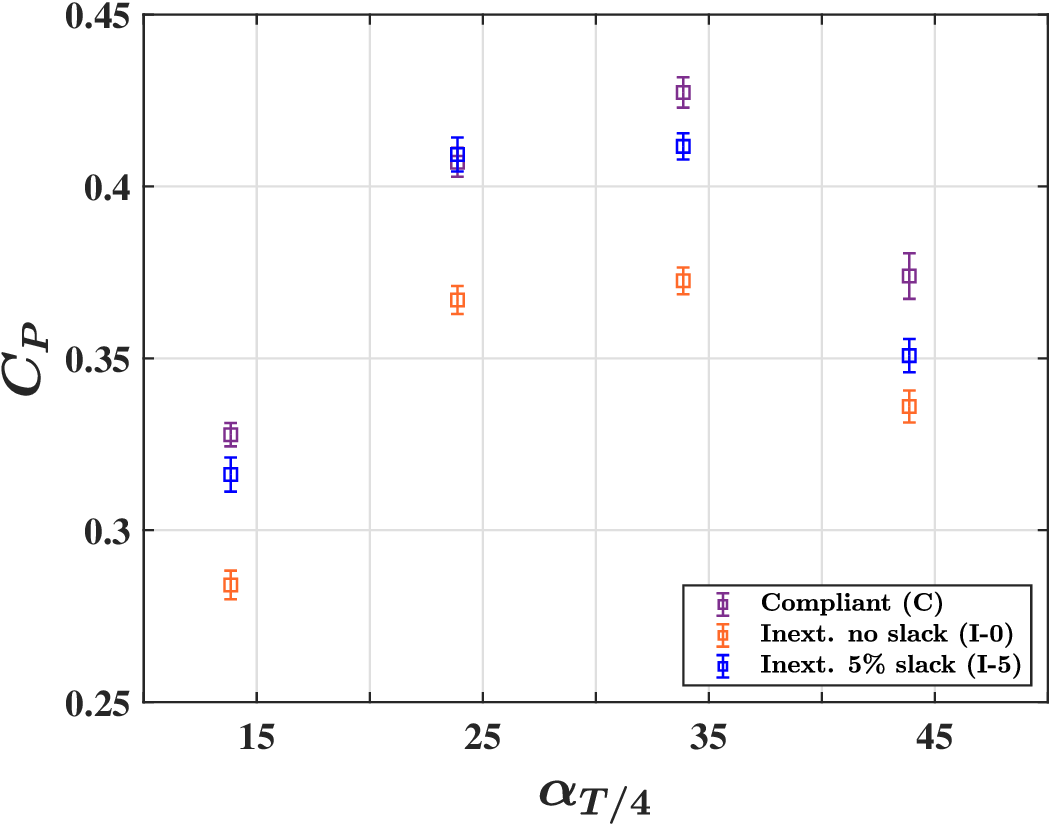}}
\end{center}
\caption{Power coefficient comparison between compliant membrane (Ae $\approx$ 20), inextensible membrane without slack and inextensible membrane with 5\% slack with varying $\alpha_{T/4}$.}
\label{fig:power_cut_figure}
\end{figure}

Based on these camber measurements, we compare the performance of three foils: the compliant membrane foil (C), an inextensible membrane foil with zero slack (I-0) and an inextensible membrane foil with 5\% slack (I-5) (a comparable camber to the $\at$ = 14$^\circ$ and 24$^\circ$ cases shown in Fig.~\ref{fig:stretch}), The power coefficients for these three foils are shown in Figure~\ref{fig:power_cut_figure} over a range of $\at$. 

\begin{figure*}
\includegraphics[width = \textwidth]{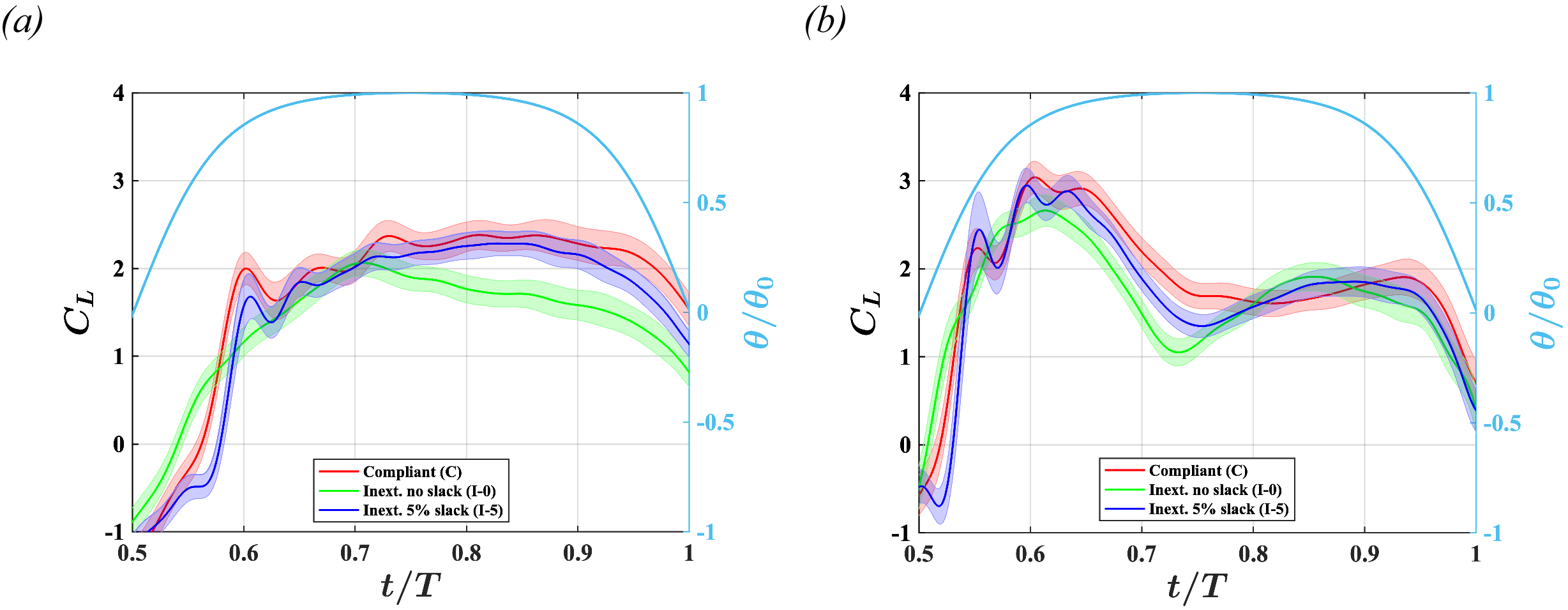}
\caption{a) Lift profiles of compliant and inextensible membranes at $\alpha_{T/4} = 14^\circ$. b) Lift profiles of compliant and inextensible membranes at $\alpha_{T/4} = 44^\circ$.}
\label{fig:lift_combined}
\end{figure*}

Both foils that allow for camber - C and I-5  - outperform the inextensible foil, I-0, over all values of $\at$. 
We observe that the performance of the C and I-5 foils are closely matched at the low $\alpha_{T/4}$ cases:  $14^\circ$ and $24^\circ$. These conditions correspond to the cases for which the mean stretch of the C foil is close to the prescribed 5\% slack of the I-5 foil (Figure ~\ref{fig:lift_combined}). At the higher values of $\alpha_{T/4} = 34^\circ$ and $44^\circ$, the compliant foil significantly outperforms the I-5 foil, and achieves values of $\lambda$ ranging between 1.1 and 1.15, larger than that prescribed of the I-5 foil. 

This improved performance can be understood by comparing the lift-vs-time of these three wings at both low and high values of $\at$ (Figure~\ref{fig:lift_combined}).
For the case of $\at = 14^\circ$, we notice that for the majority of the cycle the C and I-5 wings display closely matched lift profiles despite the elastic oscillations of the C foil observed in the stretch measurements. Both foils exhibit a rapid rise in lift ($t/T \sim 0.6$) caused by the growth of the LEV and, at this low value of $\at$, the vortex remains attached and $C_L$ is sustained at its peak value over the duration of the heave cycle.  


One subtle difference between the C and I-5 lift profiles is their pitch reversal response. The C wing displays a smooth transition from negative to positive lift at $t/T \approx 0.55$ associated with the gradual decambering and recambering of the wing as it goes from negative to positive angle of attack. In contrast, the lift profile of the I-5 wing exhibits a ``stutter'' -  a delayed lift response followed by a rapid reversal which is associated with the delay of the inextensible membrane sheet which maintains a negative camber at small positive angles of attack followed by a sudden snap through to positive camber.

Switching to the lift profiles for the highest pitching amplitude, $\alpha_{T/4} = 44^\circ$, we see the same stutter in the lift of the inextensible wing as the pitch angle rises (earlier, now at $t/T \sim 0.52$ because the pitch reversal is faster).  In contrast to the lower $\at$ case, we see that all three wings begin to stall at $t/T \sim 0.65$. The inextensible membrane without slack is the first to stall followed by the inextensible wing with slack - agreeing with previous observations that uncambered and inextensible wings exhibit the sharpest stall behavior \cite{Song}.  This delayed stall, which occurs at peak pitch angle and, more importantly, at peak heave velocity contributes to the increased energy harvesting achieved by wings with camber (Figure~\ref{fig:power_cut_figure}).
The two inextensible wings also achieve lower minima in $C_L$ of 1.1 and 1.4 followed by the growth of a second peak while the compliant wing smoothly transitions to a second stable lift coefficient, $C_L \approx 1.75$.

These trends in the lift-vs-time are completely consistent with the stretch measurements (Figure~\ref{fig:stretch}).
The sharp decambering of the compliant membrane wing in the $\alpha_{T/4} = 44^\circ$ case observed in the stretch measurements (Figure~\ref{fig:stretch}) coincides with the initiation of stall at t/T = 0.65. 

In summary, at lower angles of attack, $\alpha_{T/4} = 14^\circ,~ 24^\circ$, the compliant membrane wing exhibits a roughly constant deformation similar to the shape of the inextensible membrane wing while at higher angles of attack, $\alpha_{T/4} = 34^\circ, 44^\circ$, the membrane wing exhibits increased deformation and a stabilizing feedback behavior. 


\section{Conclusions}
The kinematic parameter space of the oscillating foil turbine was studied utilizing a compliant membrane foil. The optimum efficiency was found to occur at a lower reduced frequency than previously reported for rigid hydrofoils, $f^* = 0.11$ and $0.15$ respectively. Given Kinsey and Dumas's findings that optimal kinematics are associated with vortex shedding at the pitch reversal, we expect increased leading edge vortex stability to result in later shedding and a lower optimal oscillation frequency~\cite{Kinsey2008,Kinsey2010,kim2017}. 

In order to separate the roles which elasticity and dynamic shape-morphing play in leading-edge vortex stability on membrane wings, the energy harvesting performance of a compliant membrane wing, an inextensible membrane with slack (s = 1.05) and an inextensible membrane without slack (s = 1.00) were compared. Two distinct regimes of operation of the membrane foil OFTs were identified, the constant and dynamic camber regimes. In the constant camber regime which occurred for $\alpha_{T/4} \leq 24^\circ$, the compliant membrane achieved an equilibrium deformation quickly following pitch reversal. The performance of the inextensible membrane with similar slack achieved similar overall lift and power performance, although in this regime, the compliant membrane exhibited a smoother lift transition at the pitch reversal. For $\alpha_{T/4} \geq 34^\circ$, the compliant membrane exhibited increased deformation and decambered in response to stall, resulting in a softer stall than the inextensible membrane with slack.  

Thus, at low angles of attack, the LEV stability is primarily due to camber and the elasticity of the compliant membrane does not appear to present a significant benefit over the inextensible wing with slack. In contrast, at higher angles of attack the elasticity of the compliant membrane may enhance the LEV stability by decambering in response to stall. This later regime is critical to take into consideration as the optimal kinematics for this turbine occur exactly at this range of $\alpha_{T/4} = 30^\circ$ to $45^\circ$.

While the decambering of the compliant wing resulted in a softer stall behavior than the inextensible wing with a slack ratio of 1.05, other slack ratios could potentially perform better. Since increased $\alpha_{T/4}$ enlarges the gap between the feeding shear layer and the top of the wing, a larger wing camber may be necessary to stabilize the leading edge vortex. As the membrane stretch measurements indicate, the deformation of the compliant membrane wing increases with $\alpha_{T/4}$. The larger camber of the compliant wing at these higher $\alpha_{T/4}$ reduces the gap between the wing and the separation shear layer which may also play a role in the compliant wing's improved performance at large $\alpha_{T/4}$. 

Since the camber of the compliant membrane wing exceeded the stretch of the inextensible wing it is challenging to draw a direct comparison in this case. Further study into inextensible wings with larger slack ratios is necessary to better understand the LEV dynamics in this regime. Our results indicate that the optimal efficiency of the compliant membrane OFT is comparable to that which has been previously observed for a rigid membrane OFT. However, the use of even more compliant membrane foils with larger deformation may yield higher efficiencies.

\section*{Acknowledgements}
This work was supported by the National Science Foundation, CBET Award 1921359, and the Halpin award (to IU) from Brown University.

\bibliography{refs}

\end{document}